\documentclass[superscriptaddress,twocolumn,pre]{revtex4}
\usepackage{psfrag,epsfig,amsfonts,amssymb,amsmath,graphics}

\DeclareGraphicsExtensions{.eps,.ps}

\newcommand{\dir}{./}

\begin{document}


\title{Two-State Migration of DNA in a Structured Microchannel}
\author{Martin Streek}
\author{Friederike Schmid}
\affiliation{
Kondensierte Materie, Universit\"at Bielefeld, Fakult\"at f\"ur
Physik, D-33615 Bielefeld, Germany 
}
\author{Thanh Tu Duong}
\author{Dario Anselmetti}
\author{Alexandra Ros}
\affiliation{
Experimentelle Biophysik \& Angewandte Nanowissenschaften,
Universit\"at Bielefeld, Fakult\"at f\"ur Physik, 
D-33615 Bielefeld, Germany 
}
\date{\today}



\begin{abstract}

DNA migration in topologically structured microchannels with  
periodic cavities is investigated experimentally and
with Brownian dynamics simulations of a simple bead-spring
model. The results are in very good agreement with one
another. In particular, the experimentally observed migration order of
$\lambda$- and T2-DNA molecules is reproduced by the simulations. The 
simulation data indicate that the mobility may depend on the 
chain length in a nonmonotonic way at high electric fields.
This is found to be the signature of a nonequilibrium phase 
transition between two different migration states, a slow one and a
fast one, which can also be observed experimentally under appropriate
conditions. 

\end{abstract}


\maketitle


\section{Introduction}
\label{chapter+introduction}

DNA electrophoresis is one of the main techniques to separate DNA
molecules by size~\cite{Viovy2000}. Since the mobility of DNA
molecules 
does not depend on the chain length in free solution, electrophoresis 
is usually performed in gels. Given the progress of microtechnology,
much effort has been spent on the integration of DNA electrophoresis
into microfluidic devices (lab-on-a-chip devices)~\cite{Effenhauser97} and
separation strategies which rely on well-defined microscopic
structures have been explored. Various such gel-free devices
exploiting both 
DC~\cite{Schmalzing97,Turner98,Chou99,Han2000,Han2003,Duong2003} and 
AC~\cite{Bakajin2001,Huang2002-2} fields have been proposed. The
mobility of the DNA in these microstructures 
is determined by a subtle interplay of the characteristic dimensions
of the microstructure and the molecular size. Microstructured systems allow for fast
separation of kbp size DNA fragments in time scales less than one 
minute~\cite{Bakajin2001,Huang2002-2,Ros2004}.

\begin{figure}[t]
  \begin{center}
    \includegraphics[scale=0.30]{\dir/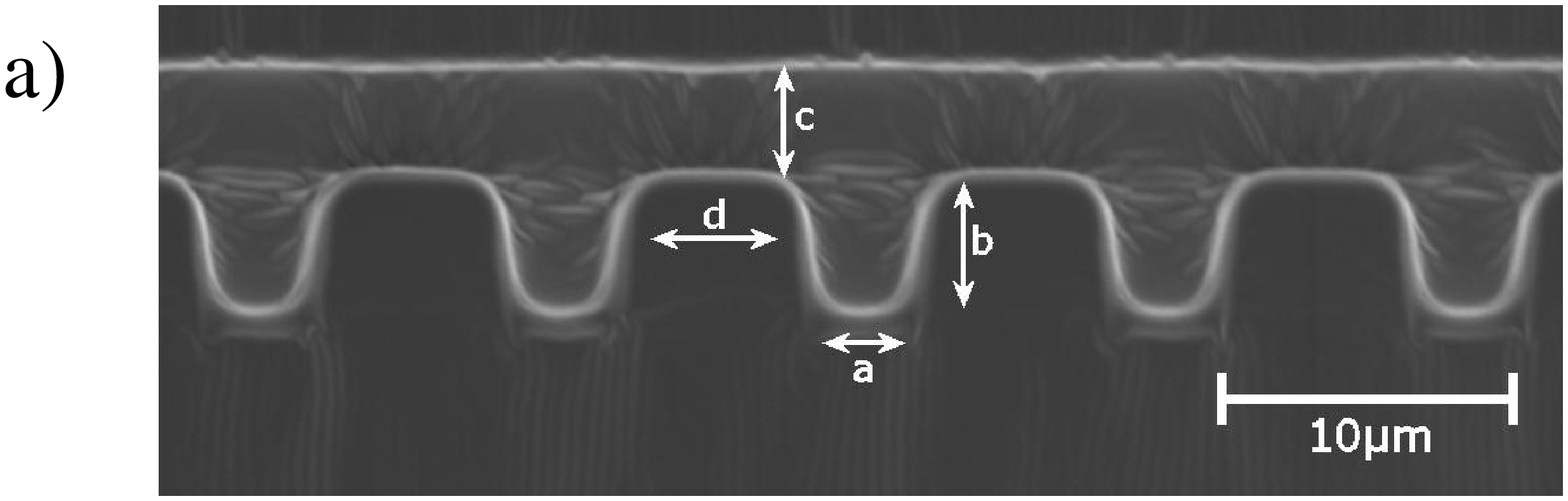}
    \includegraphics[scale=0.22]{\dir/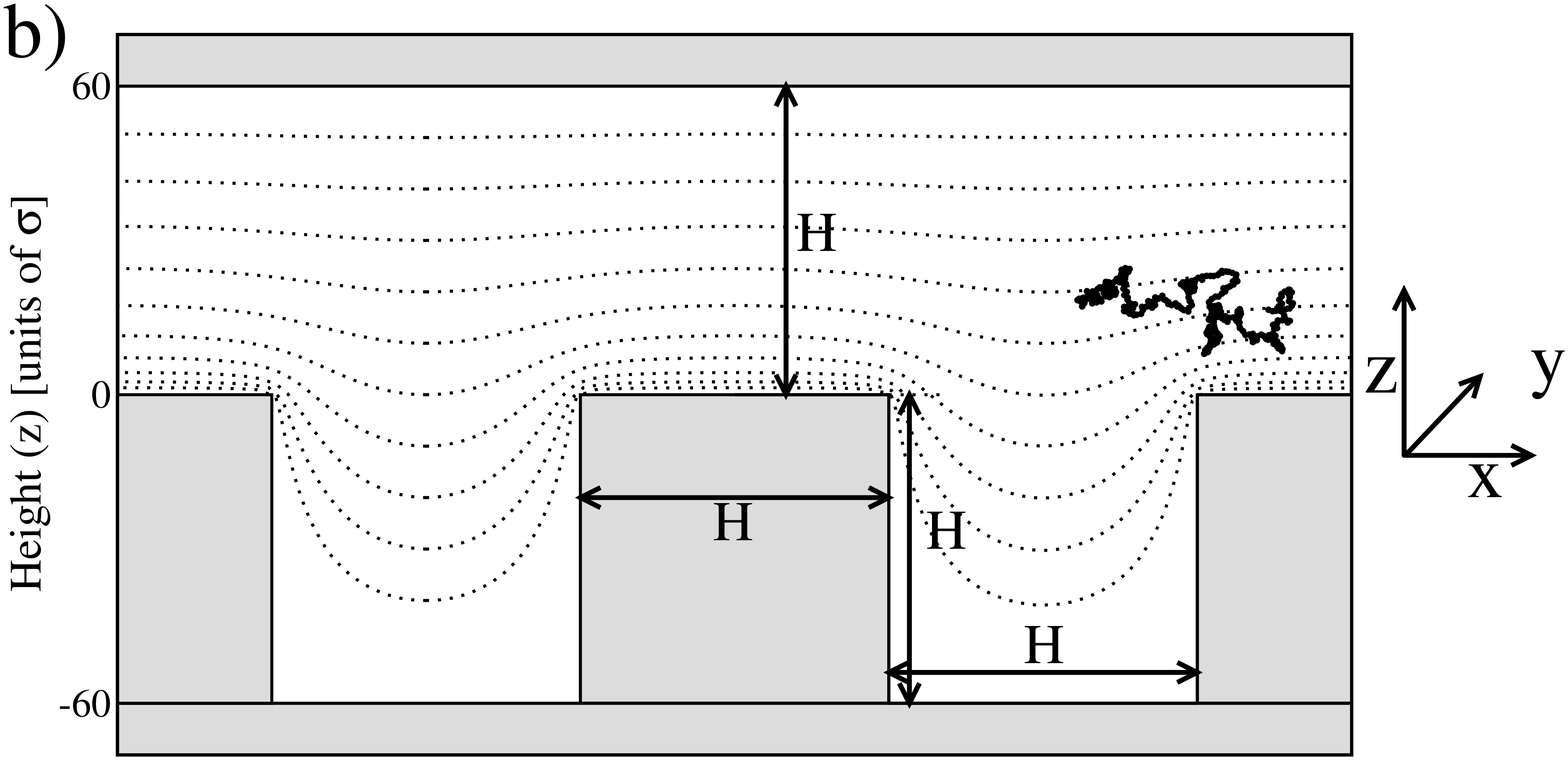}\\
  \end{center}
  \caption{
    (a) SEM image of an experimental device presented by Duong et~al.~\cite{Duong2003}, and (b) schematic
    drawing of the channels. As all size parameters are equal, only one 
    parameter $H$ is necessary to describe the structure. The
    coordinate system used in the simulations is given on the right. 
    The dotted lines represent the electric field lines. Also shown in
    (b) is a configuration snapshot from a simulation of a chain 
    with $N=320$ monomers. 
  }
  \label{fig+introduction+channel}
\end{figure}

In the present work, we focus on a geometry which has been
investigated by Han and Craighead et
al.~\cite{Han1999,Han2000,Han2002} and later by us (Duong et
al.~\cite{Duong2003}). The DNA is driven through a periodic sequence
of cavities and constrictions
(Fig.~\ref{fig+introduction+channel}) with an electrical DC field. 
The separation experiments of Han and Craighead motivated several
computer 
simulation studies~\cite{Tessier2002,Chen2003,Streek2004}, which 
reproduced the results and lead to an improved understanding of the
DNA separation mechanisms in these channels. Originally~\cite{Bakajin2001,Han1999}, the width of the
constriction was chosen of the order of $0.1 \mu\mbox{m}$ (corresponding to a few
persistence lengths of DNA molecules), and entropic trapping at the entrances of the constrictions was
proposed to be the main mechanism leading to size dependent migration of DNA in these channels.
Recently, an additional effect
that relies on diffusion has been revealed in such geometries~\cite{Streek2004}. This
second mechanism does not premise extremely narrow
constrictions. Therefore, one should also achieve DNA separation with
wider channels. 

Based on that prediction, we have investigated experimentally the
size-dependent electrophoresis in structured microchannels
(Fig.~\ref{fig+introduction+channel}) with constriction widths 
corresponding roughly to the radius of gyration of the tested DNA
molecules~\cite{Duong2003,Ros2004}. We found that separation by size 
is possible~\cite{Ros2004}. Unexpectedly,
however, the relation between mobility and 
chain length turned out to be inverse to that observed by Han and
Craighead: the short $\lambda$-DNA (48 kbp) migrated faster than the
longer T2-DNA (164 kbp)! 

In the present paper, we explore the reasons for this unforeseen behavior 
by Brownian dynamics simulations of a simple bead-spring 
model. Models of this kind have been used successfully in the past to
study the migration of DNA in various
geometries~\cite{Deutsch87,
  Deutsch88,Matsumoto94,Noguchi2001,Streek2004}. 
Our simulations reveal a surprisingly complex phenomenology. If the 
model parameters are adjusted to those of the experiment, we reproduce
the experimental findings. With other model parameters (smaller
structures, lower electric field), we recover the originally expected
increase of mobility with increasing chain length. As we show in this
paper, the reversal of this trend turns out to be the signature of a
nonequilibrium first order phase transition, which emerges at even
higher fields or in even larger structures. However unfavourable for separation in specified geometries, this
prediction implies the existence of two migration states, which could indeed be observed in migration
experiments. 

The paper is organized as follows: the simulation model and the
simulation technique are described in Sec.~\ref{chapter+model}, and
the experimental setup in Sec.~\ref{chapter+experimental-setup}. In
Sec.~\ref{chapter+simulation-mobilities}, we briefly discuss the
simulations corresponding to the earlier experimental study by Duong
et al.~\cite{Duong2003}. The results on the two coexisting migration
states are presented in Sec.~\ref{chapter+phase-transition}, and the
possible transition mechanisms are discussed in
Sec.~\ref{chapter+transition-mechanism}. We summarize and conclude in
Sec.~\ref{chapter+conclusions}. 



\section{The Simulation Model}
\label{chapter+model}


\subsection{Definition of the model}
\label{chapter+model-general}

We describe the DNA on a coarse-grained level by a bead-spring model. 
Several base pairs of DNA are represented by one bead or ``monomer''. 
Neighboring beads are connected by harmonic springs 
\begin{equation}
\label{eqn+model+spring}
V_{\mbox{\tiny sp}}(r) = \frac{k}{2} r^2,
\end{equation}
where $r$ is the distance between the two neighboring monomers. 
All beads interact with a purely repulsive 
Weeks-Chandler-Andersen (WCA) potential
\begin{equation} 
\label{eqn+model+lj-potential}
V_{\mbox{\tiny pair}}(r)/k_B T = \left\{
\begin{array}{l@{\quad:\quad}l} 
(\frac{\sigma}{r})^{12} - (\frac{\sigma}{r})^6 +
    \frac{1}{4} & (\frac{r}{\sigma}) \le 2^{1/6},
    \\ 0 & \mbox{otherwise}, 
  \end{array}
\right.  
\end{equation}
where $r$ is the distance between two monomers and $\sigma \equiv 1$ sets
the range of the interaction. 

The walls are assumed to be soft and purely repulsive. The potential
describing the interaction with the walls is the same as that acting
between the monomers (Eq.~\ref{eqn+model+lj-potential}, with $r$ as
the the minimum distance between a bead and the wall). In the
corners, the forces of the adjacent walls are summed up. To model the
finite depth of the device, we also introduce walls in $y$-direction. 

Each bead carries a charge $q$ and is subject to an electrical force
$\vec{f}_{\mbox{\tiny el}} = q \vec{E} = -q \nabla \Phi$, but charges
do not interact with one another in our model. The field is calculated
by solving the Laplace equation ($\Delta \Phi = 0$) in the channel. We
use von Neumann boundary conditions at the walls ($\vec{n} \cdot
\nabla \phi = 0$, with $\vec{n}$ the surface normal) and impose a constant potential difference between the
inlets of the device. A numerical solution $\Phi(\vec{r})$ is obtained from the software
program ``Matlab'' (Mathworks, US), using a finite element solver. The grid values of the
derivatives are interpolated bilinearly. The field lines
are shown in Fig.~\ref{fig+introduction+channel}. Throughout the rest
of this paper, the field denoted by $E$ refers to the total potential
difference divided by the total channel length, and can thus be
regarded as the characteristic strength of the electric field.  

The motion of the chain is described by Langevin dynamics, {\em i.e.}
the solvent surrounding the chain is replaced by a Brownian force
$\vec{\eta}_i$ and an effective friction coefficient $\zeta \equiv 1$,
which fulfill the conditions~\cite{Doi86} 
\begin{eqnarray}
  \label{eqn+model+random-force-a}
  <\eta_i> & = & 0 \\
  \label{eqn+model+random-force-b}
  <\eta_{i,\alpha}(t) \eta_{j,\beta}(t')> &=& 
2 \zeta \: k_B T \: \delta_{ij} \delta_{\alpha\beta} \: \delta(t-t'), 
\end{eqnarray}
with monomer indices $i, \; j = 1 \dots N$, cartesian directions
$\alpha,\beta \in \{x,y,z\}$ and $t$, $t'$ two given times. The
equations of motion are thus
\begin{eqnarray}
  \label{eqn+model+langevin-verlet-dynamics-a}
  \dot{\vec{r}}_i &=& \vec{v}_i \\
  \label{eqn+model+langevin-verlet-dynamics-b}
  m\dot{\vec{v}}_i &=& \vec{f}_i - \zeta \vec{v}_i + \vec{\eta}_i, 
\end{eqnarray}
where $\{\vec{r}_i\}$ and $\{\vec{v}_i\}$ represent the locations and
the velocities of the monomers, respectively. The vectors
$\{\vec{f}_i\}$ sum up forces acting on a monomer (harmonic
spring, repulsive monomer interaction, wall interaction, external
electric field). Here, each bead carries a mass $m$, but simulations
with vanishing mass were also carried out for comparison (see
Sec.~\ref{chapter+transition-mechanism}). With $m = 0$, the equations of
motion reduce to 
\begin{equation}
  \label{eqn+model+langevin-no-verlet-dynamics}
  \zeta\dot{\vec{r}}_i = \vec{f}_i + \vec{\eta}_i.
\end{equation}

The natural units of our simulation are defined in terms of the bead
size $\sigma$, the friction coefficient $\zeta$, the bead charge
$|q|$, and the temperature $T$. Based on these quantities, the energy
unit is $k_{\mbox{\tiny B}} T$, the length unit is $\sigma$,
the time unit is $t_0 = \zeta\sigma^2/k_{\mbox{\tiny B}} T$, and the
electric field unit is $E_0 = k_{\mbox{\tiny B}} T / \sigma |q|$. 

The dynamical equations (Eq.~\ref{eqn+model+langevin-verlet-dynamics-a}
and Eq.~\ref{eqn+model+langevin-verlet-dynamics-b}) are integrated
with a Verlet algorithm using a timestep $\Delta_{\mbox{\tiny t}} =
10^{-2}t_0$. In the case of the relaxational dynamics in
Eq.~\ref{eqn+model+langevin-no-verlet-dynamics}, we use an Euler 
algorithm with the time step $\Delta_{\mbox{\tiny t}} =
10^{-4}t_0$. The stochastic noise $\vec{\eta}_i$ was implemented by
picking random numbers at every 
time step. As shown in Ref.~\onlinecite{Duenweg91}, the
random numbers do not have to be Gaussian distributed, as long as they
fulfill Eq.~\ref{eqn+model+random-force-a} and
Eq.~\ref{eqn+model+random-force-b}. Those used in our simulation were 
evenly distributed inside the unit sphere. Unless stated otherwise,
the run lengths were $4 - 20 \cdot 10^8 \Delta_{\mbox{\tiny t}}$ with
equilibration times of $2 \cdot 10^7 \Delta_{\mbox{\tiny t}}$. 

The spring constant in
Eq.~\ref{eqn+model+spring} was chosen very high, $k = 100
k_{\mbox{\tiny B}} T / \sigma^2$. This ensures that no chain crossing occurs~\cite{Streek2002}. The
equilibrium length of each bond is $0.847\sigma$. 

The static properties of free chains of length $N$ correspond to 
those of self-avoiding random walks~\cite{Streek2002} with the
persistence length $l_{\mbox{\tiny p}} = 1.6 \sigma$ and the radius of 
gyration $R_{\mbox{\tiny g}} \approx 0.5 \sigma \cdot N^{\nu}$. 
Here $\nu=0.588$ is the Flory exponent. The dynamical properties are
characterized by the diffusion constant $D=k_{\mbox{\tiny B}}T/N \zeta
= N^{-1} \sigma^2/t_0$, the mobility $\mu_0 = |q|/\zeta = \sigma/E_0
t_0$, which is independent of the chain length $N$, and the decay time
of the drift velocity $\tau_0 = 
m/\zeta$~(Eq.~\ref{eqn+model+langevin-verlet-dynamics-b}). The
quantity $\tau_0$ sets the time scale on which inertia effects are
significant. In most simulations presented here, it was set to $\tau_0
= t_0$ ({\em i.e.} $m = \zeta t_0$), based on the assumption that the
relevant time scales for the dynamical processes of interest here are
much longer. In some cases, however, simulations with vanishing
$\tau_0$ (vanishing mass) were also necessary (see
Sec.~\ref{chapter+transition-mechanism}). On time scales larger than
$\tau_0$, the chains behave like standard (self-avoiding) Rouse
chains~\cite{Doi86} with the Rouse relaxation time
$\tau_{\mbox{\tiny R}} \approx 0.045\: t_0 \cdot N^{1+2 \nu}$. 



\subsection{Correspondence between the model and the experiment}
\label{chapter+model+adaptation}

In the experiments, the mobility of $\lambda$-DNA (48 kbp) and
T2-DNA (164 kbp) in microchannels with geometries such as sketched
in Fig.~\ref{fig+introduction+channel} was studied. Details are given
in Ref.~\onlinecite{Duong2003} and
Sec.~\ref{chapter+experimental-setup}. We will now discuss how the
parameters and units of our model can be related to those of the
experiment. 

The energy unit is simply given by the temperature, $k_{\mbox{\tiny
    B}} T = 300 k_{\mbox{\tiny B }} \mbox{K} \approx 0.026 \mbox{eV}$,
    at which the experiments were carried out. Next we 
adjust the value of the length unit. In our earlier work on entropic
traps~\cite{Streek2004}, we had matched the persistence lengths of the
model chains with that of DNA. In the present case, however, the
persistence length is not an experimentally relevant length scale,
because all dimensions of the microchannels are on the order of
magnitude of the radius of gyration. Thus we can determine the length
unit by comparing the characteristic length $H$ in our model channel
(see Fig.~\ref{fig+introduction+channel}) to the corresponding
experimental value.  Note that this implies that we can relate
simulations with {\em one} channel geometry to all three experimental
channels by scaling our results accordingly. 

More specifically, the simulations were carried out using channels
with $H = 60 \sigma$. We illustrate our adaptation with the example of
the $5 \mu \mbox{m}$ microchannels. It is carried out in four
steps. First, we compare the channel height $H= 5 \mu \mbox{m} \equiv
60 \sigma$. This gives the length unit $1\sigma \equiv 83
\mbox{nm}$. Second, comparing the radii of gyration of our model
chains with that of $\lambda$-DNA,  we find that $\lambda$-DNA is
represented by chains of $N = 140$ beads, 
or 1 bead $\equiv$ 340 bp. Third, the time unit is adjusted by
matching the diffusion constant $D$ for $\lambda$-DNA. Experimentally,
Smith et.~al.~have reported $D = 0.47 \pm 0.03 \mu
\mbox{m}^2/\mbox{sec}$~\cite{Smith95}.
Comparing this with the theoretical value, $D = \sigma^2/N t_0$, we
obtain $t_0 \equiv 1.1 \cdot 10^{-4} \mbox{s}$. Hence, one second
corresponds to $9.5 \cdot 10^3 t_0$ (or $ \sim 10^6$ time
steps). Finally, the electric field unit $E_0$ is adjusted by
adjusting the mobility in large unstructured microchannels, which is chain length independent and given
by $\mu_0 = 1.84 \cdot 10^{-4} \mbox{cm}^2/\mbox{Vs}$~\cite{Duong2003}. It is
the sum of the free flow mobility $\mu_0$ and the electroosmotic
mobility $\mu_{\mbox{\tiny eof}} = (2.9 \pm 0.6) \cdot 10^{-4}
\mbox{cm}^2/\mbox{Vs}$ in our channels (see below). 
This leads to $1 \mbox{V/cm} \equiv 2.3 \cdot 10 ^{-3} E_0$. These values as well 
as those obtained for channels with $H =3\mu \mbox{m}$ and $H = 1.5\mu 
\mbox{m}$ are summarized in table~\ref{table+model+adaptation}. Unless stated
otherwise, all adaptations given in this paper refer to the $5 \mu
\mbox{m}$ channels, except in Sec.~\ref{chapter+simulation-mobilities}. 

\begin{table}[b]
  \begin{tabular}{|r||r|r|r|} \hline
    & $H = 1.5 \mu \mbox{m}$ & $H = 3 \mu \mbox{m}$ & $H = 
    5 \mu \mbox{m}$      \\\hline\hline 
    $1$ bead      & 48 bp                 & 150 bp                &
    340 bp                \\ \hline 
    $\lambda$-DNA & 1000 beads            & 330 beads             &
    140 beads             \\ \hline 
    T2-DNA      & 3500 beads            & 1200 beads            &
    490 beads             \\ \hline 
    $1 \mu \mbox{m}$     & $40 \sigma$             & $20 \sigma$
    & $12 \sigma$             \\ \hline 
    $1 \mbox{s}$  & $ 7.5 \cdot 10^5 t_0$      & $6.2 \cdot 10^4 t_0$
    & $9.5 \cdot 10^3 t_0$\\ \hline 
    $1 \mbox{V/cm}$      & $9.8 \cdot 10^{-5} E_0$ & $5.9 \cdot
    10^{-4} E_0$ 
    & $2.3 \cdot 10^{-3} E_0$ \\ \hline 
    $E_0$         & $10 \mbox{kV/cm}$    & $1.7 \mbox{kV/cm}$   & $430
    \mbox{V/cm}$ \\ \hline  
  \end{tabular}
  \caption{Adaptation of the simulation units to various channel
  sizes. All values have been rounded to two digits.} 
  \label{table+model+adaptation}
\end{table}

We set the depth of the device to $60\sigma$. For comparison, we also performed
simulations with infinite depth of the device and found no qualitative
difference (data not shown). In our case, the depth has no significant
influence on the dynamics. 

The model disregards a number of important physical effects. 

First, electrostatic interactions are neglected. The Debye screening length
of DNA in typical buffer solutions is about $2 \mbox{nm}$, which is
comparable to the diameter of the polymer and much smaller than the
persistence length of DNA (roughly $50 \mbox{nm}$). 

Second, PDMS exhibits silanol groups on its surface which dissociate 
under the experimental conditions. Thus the experiments were carried
out in microchannels with negatively charged surfaces. This implies
the generation of cathodic electroosmotic flow, and the resulting
mobility of DNA molecules is a sum of the electroosmotic and
electrophoretic mobilities. The DNA molecules migrate to the anode,
indicating that electrophoresis overcomes electroosmosis. In the
simulation, we do not incorporate any electroosmotic flow, or flow in
general. However, recent reports~\cite{Cummings2000,Santiago2001} show that for sufficiently low Reynolds
number, the flow outside the Debye layer at the walls is rotation free and proportional to the electric field
with a fixed proportionality constant $\alpha$. Let $\vec{v}_0(\vec{r})$ denote
the flow field at a given position $\vec{r}$ and $\vec{f}_{\mbox{\tiny o,}i}$ the forces acting on monomer
$i$, excluding the electric force $\vec{f}_{\mbox{\tiny el,}i} = - q \cdot \nabla \Phi(\vec{r}_i)$. Using
$\vec{v}_0(\vec{r}) = \alpha \cdot \nabla \Phi(\vec{r})$, the equation of motion
(Eq.~\ref{eqn+model+langevin-verlet-dynamics-b}) now becomes 
\begin{equation}
  \begin{array}{*{3}{rcl}}
    m \dot{\vec{v}}_i & = & \vec{f}_{\mbox{\tiny o,}i} + \vec{f}_{\mbox{\tiny el,}i} - \zeta (\vec{v}_i +
    \vec{v}_0(\vec{r})) + \eta_i \\
    & = & \vec{f}_{\mbox{\tiny o,}i} - ( q + \zeta \alpha) \cdot \nabla \Phi(\vec{r}) + \eta_i,
  \end{array}
\end{equation}
and for the overdamped dynamics (Eq.~\ref{eqn+model+langevin-no-verlet-dynamics}) one finds 
\begin{equation}
  \zeta \dot{\vec{r}}_i = \vec{f}_{\mbox{\tiny o,}i} - ( q + \zeta \alpha) \cdot \nabla \Phi(\vec{r}) +
  \eta_i. 
\end{equation}
These equations are formally equivalent to Eq.~\ref{eqn+model+langevin-verlet-dynamics-b} and
Eq.~\ref{eqn+model+langevin-no-verlet-dynamics} with a rescaled charge $q$. 
The Reynolds number for the electroosmotic flow is easily found for our device. Using the viscosity of
water, the electroosmotic mobility given above and a characteristic length of $5\mu\mbox{m}$, the Reynolds
number is roughly $10^{-3}$ for electric fields 
of 100~V/cm in our channels. This justifies our adaptation to the overall mobility given above. 

Third, hydrodynamic effects are not taken into account. This
approximation must be questioned. On the one hand, DNA is always
surrounded by counter ions, which are dragged into the opposite
direction of the DNA. Thus the DNA molecule experiences not only
hydrodynamic drag, but also an extra friction from the solvent
molecules. In free solution, these two effects cancel each
other~\cite{Viovy2000}. This ``hydrodynamic screening'' accounts for
the free-draining property of DNA. However, the total
cancellation fails if the DNA molecule is blocked by a geometric
barrier~\cite{Long96}. In that case, the counter ions will not be
immobilized, since counterions are still free to move. Furthermore,
hydrodynamic interactions affect the diffusion constant $D$. In our
model, it scales the chain length $N$ like a Rouse chain ($D \propto 
1/N$). Including hydrodynamic interactions, one would expect Zimm
scaling ($D \propto 1/R_g \propto 1/N^{\nu}$). In experiments, the
diffusion constant of DNA is found to scale as
$D \propto 1/N^{0.672}$~\cite{Stellwagen2003}.  

Unfortunately, a full simulation which treats hydrodynamic
interactions correctly, takes into account the dynamical
counter ion distribution, and includes electrostatic interactions, is
a formidable task and impossible with standard computer resources. The
simplifications described above will affect the quantitative results,
but presumably neither by orders of magnitude nor qualitatively. 




\section{Experimental Setup}
\label{chapter+experimental-setup}
The fabrication method used for the structured microchannels is based
on soft lithography of poly(dimethylsiloxane)(PDMS) and is described
in detail in Ref.~\onlinecite{Duong2003}. Briefly, a photoresist (SU-8
from Microresist, GER) coated Si-wafer (CrysTec, GER) is exposed
through a chromium mask (DeltaMask, NL). After developing and
hard-baking liquid PDMS (Sylgard 184, Dow Corning, USA) is poured onto
the master containing the inverted microstructure and baked for 3 h at
$75^\circ\mbox{C}$. Peeling off the PDMS slab, punching reservoir holes
through the PDMS and covering it with a clean microscope glass slide
results in the microchip. Fig. 2b shows a scanning electron micrograph
(SEM) image of the topview of a microchannel with H=$5\mu
\mbox{m}$. In comparison to the simulated geometry, the corners of the
cavities are rounded and the lengths a, b, c, and d in Fig.~\ref{fig+introduction+channel} are not
exactly equal. The exact values were a=$3.6\mu \mbox{m}$, b= $6.0\mu
\mbox{m}$, c=$3.7\mu \mbox{m}$ and d=$4.7\mu \mbox{m}$. The depth of
the microchannel was $2.8\mu \mbox{m}$. Other channel geometries
mentioned in this work are described in detail in
Ref.~\onlinecite{Duong2003}.  


For fluorescence imaging, $6 \, \mbox{pM} \, \lambda$- or T2-DNA
(Fluka, Germany) in the same buffer ($10 \, \mbox{mM}$ Tris at
$\mbox{pH} \, 8.3$ containing $50 \, \mbox{mM}$ NaCl, $1\, \mbox{mM}$
ethylenediaminetetraacetic acid (EDTA) and $2\%$ (v/v)
$\beta$-Mercaptoethanol) were adjusted to a YOYO-1 (Molecular Probes,
USA) in a basepair ratio of $1:7.5$. A sensitive fluorescence
video-microscopy setup is used to record DNA migration, as described
in Ref.~\onlinecite{Duong2003}. Briefly, it consists of an inverted
microscope (Axivert 100, Zeiss, GER) with a filterset for fluorescence
observation (XF100-3, Omega, USA), a 100x oil immersion objective
(Plan Neofluar NA 1.3, Zeiss, GER) and a sensitive CCD-camera (Imager
3LS, LaVision, GER). Data acquisition and automated data analysis is
performed in DaVis 5.4.4 (LaVision) using cross-correlation
analysis. Electric fields are applied with power supplies from FUG
(MCN 14-2000, GER). 

The electroosmotic mobility was determined according to the current
monitoring method~\cite{Huang88}. Linear channels of $20 \, \mu
\mbox{m}$ width and depth were filled with water immediately after
assembly, which was exchanged by the $10 \, \mbox{mM}$ sample buffer
without DNA before the experiment. The buffer in the anode reservoir
was then exchanged by $8 \, \mbox{mM}$ Tris buffer with otherwise
identical composition as the $10 \, \mbox{mM}$ buffer. A constant
electric field of $400 \,\mbox{V/cm}$ was applied and the current was
recorded until a stable lower current value was reached. The recorded
current decay times $\tau_{\mbox{\tiny eof}}$ were used to calculate
the electroosmotic mobility according to: $\mu_{\mbox{\tiny eof}} =
\frac{l}{E \tau_{\mbox{\tiny eof}}}$, with $l$ the length of the
channel, and $E$ the applied electric field. 



\section{Low electric field: steady-state migration} 
\label{chapter+simulation-mobilities}

We begin by presenting simulations for relatively low electric
fields $E$, which can be compared to the previously published
experimental results of Duong et al.~\cite{Duong2003}. In this
section, we will adjust our model units to the channels with
characteristic size $H = 3 \mu \mbox{m}$
(cf. Sec.~\ref{chapter+model+adaptation}), where Duong et~al. had
found particular clear evidence of a size-dependent mobility.

\begin{figure}[ht]
  \begin{center}
    \includegraphics[scale=0.2]{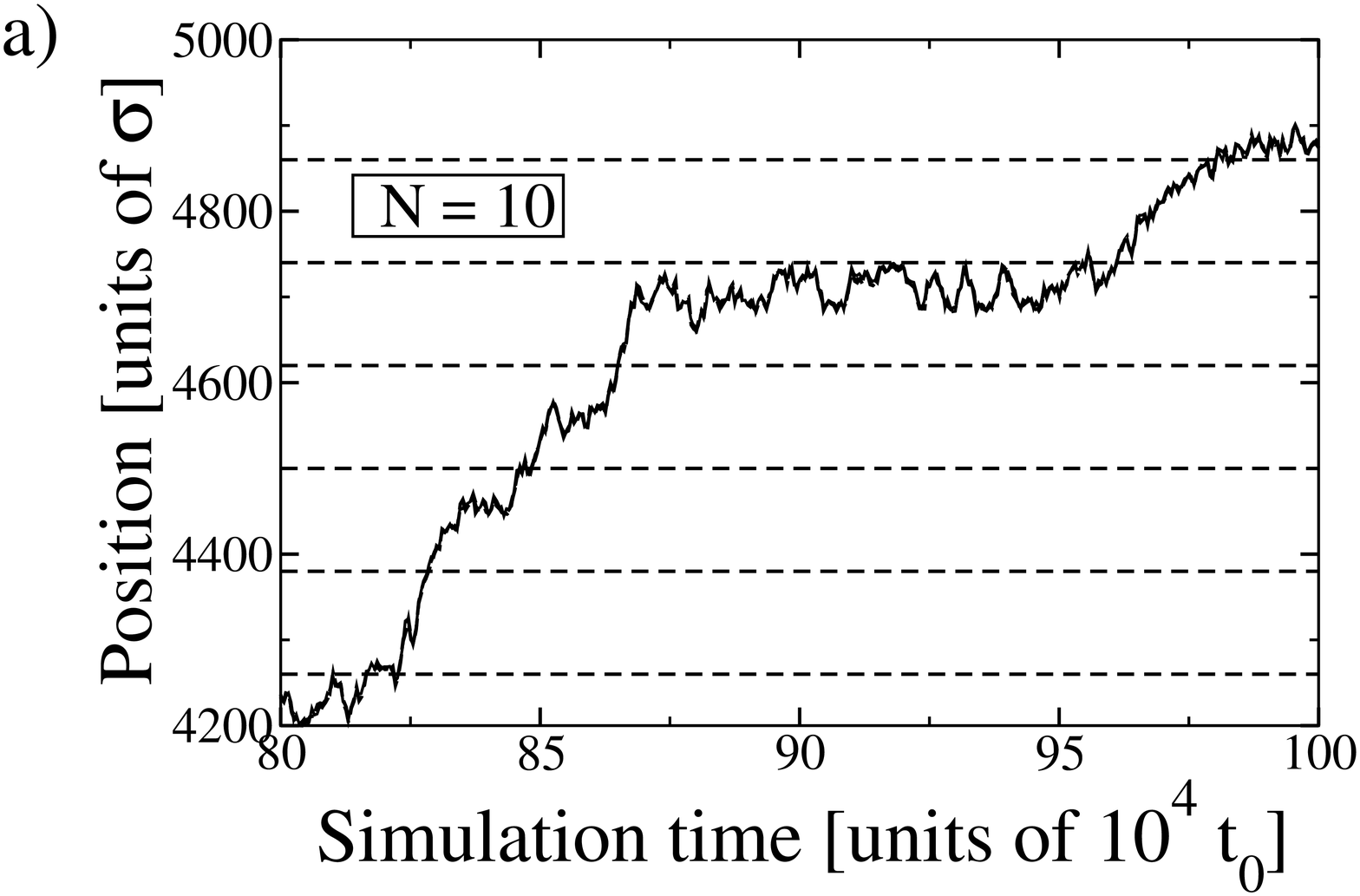}
    \includegraphics[scale=0.2]{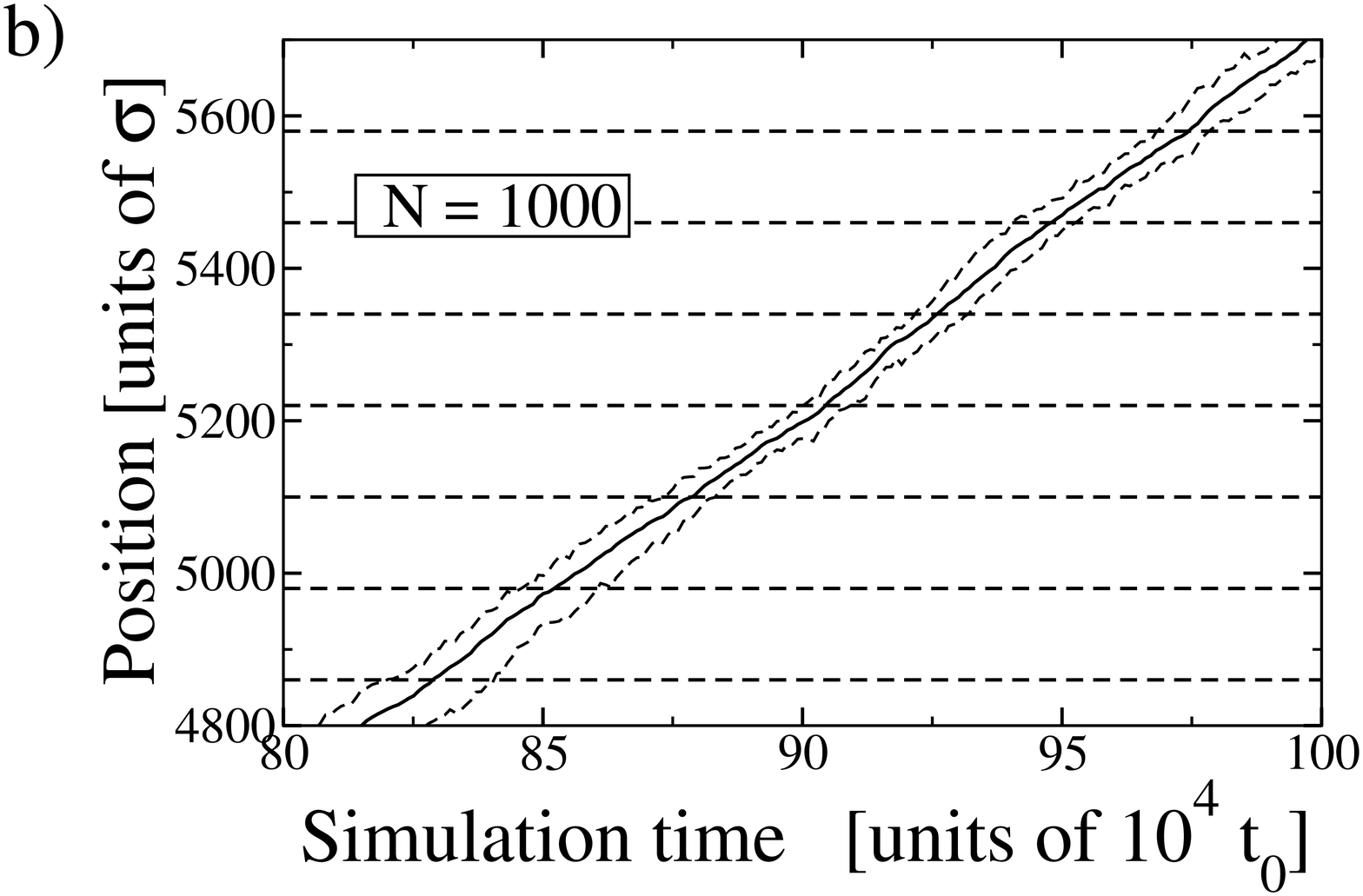}
  \end{center}
  \caption{Trajectories for chains of length (a) N = 10, and (b) N =
    1000, at $E = 0.005 E_0$. The horizontal dashed lines mark the
    beginning of the narrow regions. The solid line in the middle
    shows the position of the center of mass, the dashed lines
    indicate the leading and trailing monomers. For $N
    = 10$ (a), these lines cannot be distinguished from each other.} 
  \label{fig+mobil+trajec}
\end{figure}

\begin{figure}[ht]
  \begin{center}
    \includegraphics[scale=0.28]{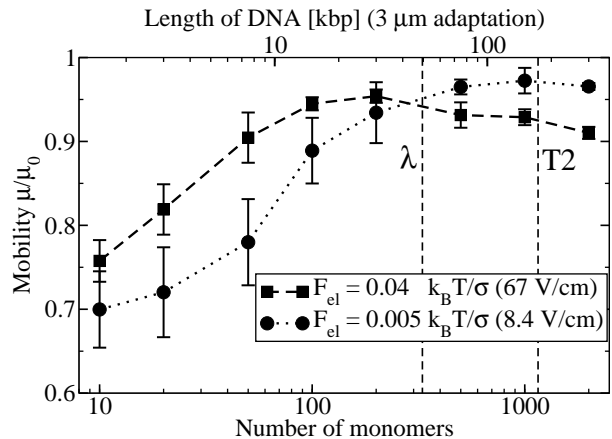}
  \end{center}
  \caption{Mobility as a function of chain length for $E = 0.005 \:
  E_0 \; (8.4 \: \mbox{V/cm})$ and $E = 0.04 \: E_0 \; (67 \: \mbox{V/cm})$. The 
  vertical dashed lines indicate the 
  length of $\lambda-$ and T2-DNA, with adaptation for the $3\mu
  \mbox{m}$ constrictions. Experimentally, Duong et.~al. have reported
  $\mu /\mu_0 = 0.99 \pm 0.1$ for $\lambda$-DNA and $\mu /\mu_0 = 0.63 \pm 0.03$ for 
  T2-DNA~\cite{Duong2003}.} 
  \label{fig+mobil+mobility}
\end{figure}

The electric field was varied from $E = 0.0025 E_0$ up to $E = 0.04
E_0$, which corresponds to experimental fields in the range of $4.2
\mbox{V/cm}$ up to $67 \mbox{V/cm}$. The chain lengths range from $N =
10$ up to $N = 2000$ monomers, modeling DNA strands from 1.5 kbp up to
290 kbp. Fig.~\ref{fig+mobil+trajec} shows typical trajectories
obtained at $E = 0.005 E_0$. The trajectories are similar to those
obtained for the entropic trap geometry~\cite{Streek2004}. For $N =
10$, the trajectory is completely dominated by diffusion. Long chains
with $N = 1000$ monomers are hardly affected by the constrictions --
much less, in fact, than in the case of the entropic traps~\cite{Streek2004}. This is
because the width of the narrow regions is of the order of the radius
of gyration; the chain is thus able to cross the constriction without
uncoiling. The trapping mechanism suggested by Han et
al.~\cite{Han2000} should not apply here.  The fact that some trapping
nevertheless occurs for short chains underlines the importance of the
effect reported by Streek et al.\cite{Streek2004}. Indeed,
Fig.~\ref{fig+mobil+mobility} shows that the mobility increases as a
function of the chain length for $E = 0.005 E_0$.  
In the experimental channel, $E=0.005 E_0$ corresponds to a real
field of $E \sim 8.4 \mbox{V/cm}$. In such low fields, Duong et
al.~\cite{Duong2003} were unable to distinguish between different
mobilities for different DNA sizes; the statistical error is large
because diffusive motion becomes significant. We note that in
simulations, we can track the DNA migration over many more cavities
than in the experiments, therefore the relative error is smaller. 

At higher fields, Duong et al.~\cite{Duong2003} reported the inverse
effect -- the mobility of T2-DNA is smaller than that of
$\lambda$-DNA. If we increase the electric field in the simulations
accordingly, we find indeed that, at $E = 0.04 E_0$, the mobility
reaches a maximum at $N = 200$ beads (29 kbp) and decreases for longer
chains (Fig.~\ref{fig+mobil+mobility}). In the parameter region
corresponding to $\lambda$-DNA and T2-DNA, smaller chains migrate
faster than longer ones, but the difference in the mobility is not as 
pronounced as in experiment. Thus our simulations reproduce the
behavior observed by Duong et al.~\cite{Duong2003} qualitatively. 

This result is gratifying, but it does not yet explain why the chain 
length dependence of the mobility is suddenly reversed. In order to
explore this question, we proceeded to investigate the migration in our
structures under more extreme conditions. In the simulations, we can
increase the field strength $E$. In the experiments, this can only be
realized up to a field of approximately $100 \mbox{V/cm}$, where the
cross correlation analysis~\cite{Duong2003} fails. However,
table~\ref{table+model+adaptation} shows that we can alternatively
work with larger structures. Therefore, we have fabricated and studied
microstructures with $H=5 \mu \mbox{m}$, and the results will be
presented in the next section. 



\section{High electric field: two migration states}
\label{chapter+phase-transition}

\begin{figure}[ht]
  \begin{center}
    \includegraphics[scale=0.2]{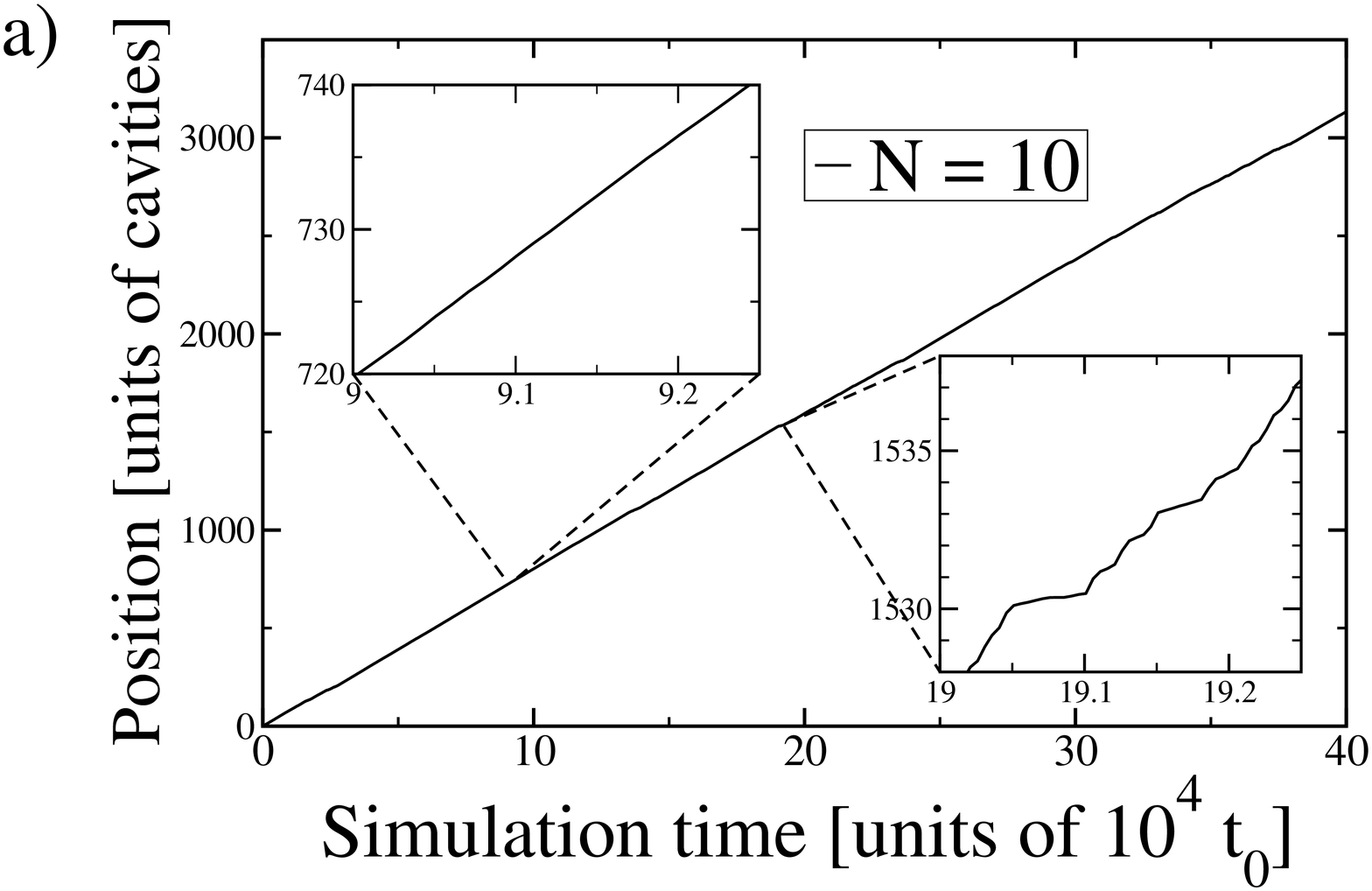}
    \includegraphics[scale=0.2]{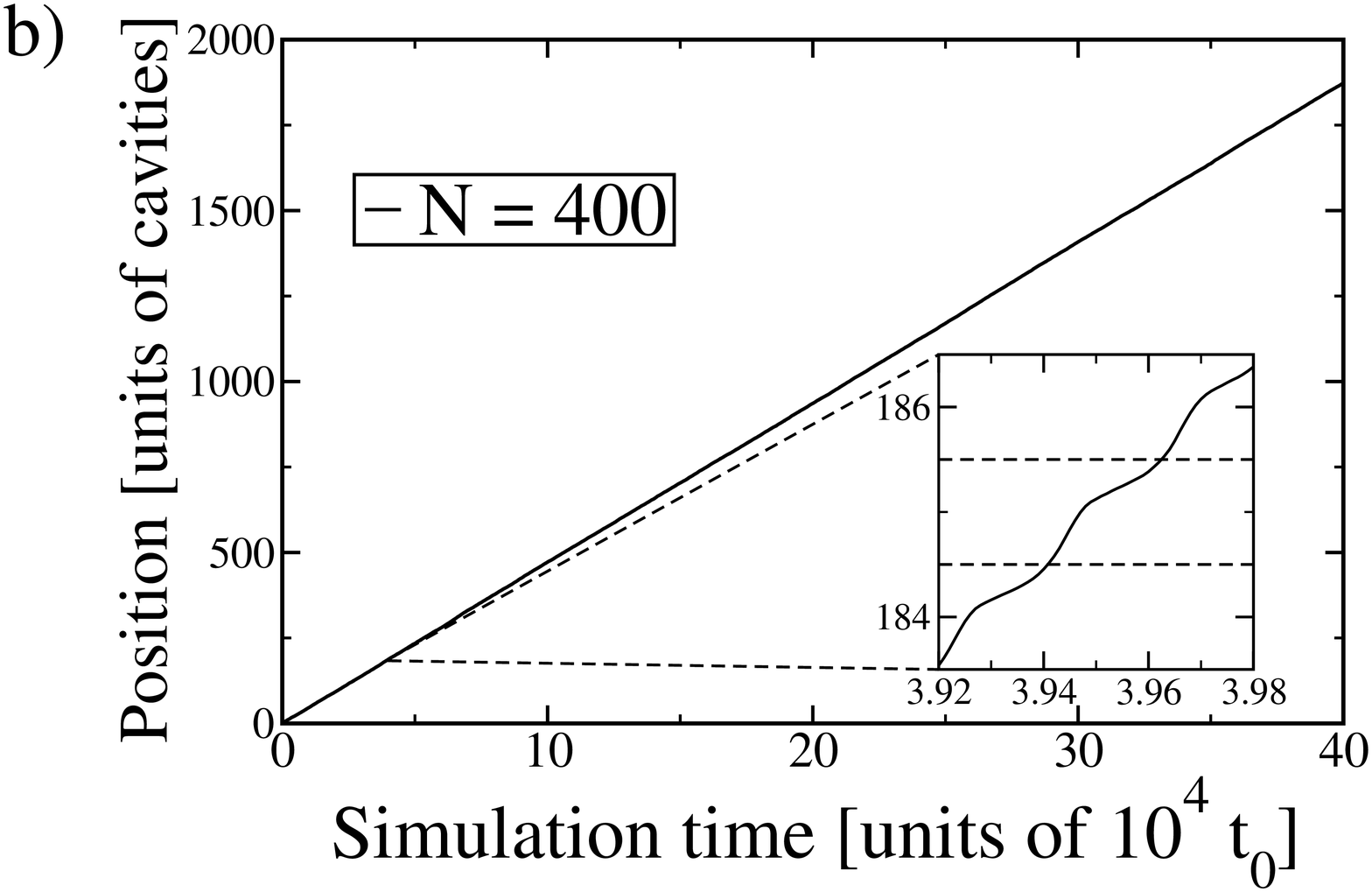}
  \end{center}
  \caption{Trajectories of a chain with a) $N = 10$ (3.4 kbp) and b)
    $N = 400$ (140 kbp) monomers at $E = E_0$ (430 V/cm). The dashed
    lines inside the inset in the $N = 400$ case represent the onset
    of the narrow regions.} 
  \label{fig+phase-transition+trajec10_400}
\end{figure}

We begin by discussing the simulation results. We started with
studying the extreme case $E = E_0$. This corresponds to 430~V/cm in
a $5 \mu \mbox{m}$ structure and is thus not accessible in
experiments. Two sample trajectories for a chain of length $N = 10$
(3.4 kbp) and $N = 400$ (140 kbp) monomers are given in
Fig.~\ref{fig+phase-transition+trajec10_400}. As can be seen from the
insets in the $N=10$ case, the chains retain some memory of their
state in the previous cavity. If a chain passes a constriction without
being trapped, it will avoid trapping at the next barrier as well. A
similar effect had already been observed in the 
entropic traps~\cite{Streek2004}. As in the simulations discussed in
the previous section, chains may still get trapped in the corners of
cavities, but at a lower rate due to the reduced diffusion time inside
a single cavity. For $N = 400$, the migration is periodically fast and
slow, indicating that the polymer penetrates the low field regions in
the cavities. Over long simulation times, both the trajectories for
$N=10$ and $N=400$ are quite regular.  

\begin{figure}[b]
  \begin{center}
    \includegraphics[scale=0.25]{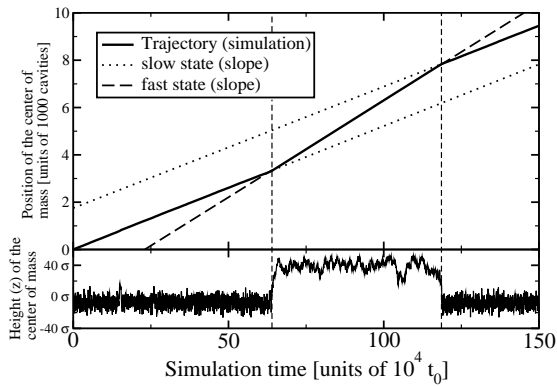}
  \end{center}
  \caption{Trajectory of a chain with $N = 200$ (70 kbp) monomers at
    $E = E_0$, along the $x$ direction (top) and the $z$ direction
    (bottom). Note the two different migration speeds which can be
    related to the penetration depth into the wide cavities (the
    dashed vertical lines are just guides to the eye).}  
  \label{fig+phase-transition+trajec200_combi}
\end{figure}

The situation is qualitatively different for $N = 200$ (70 kbp). The
trajectory is smooth on short time scales, but occasionally, the speed
of migration changes abruptly
(Fig.~\ref{fig+phase-transition+trajec200_combi}). One observes two 
states of migration, one of which is fast and the other is slow. The
lower panel of Fig.~\ref{fig+phase-transition+trajec200_combi} shows
that the two states are associated with two distinctly different
$z$-positions of the center of mass: the polymer migrates faster when
it stays in the homogeneous field in the upper part of the
channel. From Fig.~\ref{fig+phase-transition+trajec200_combi}, both
states are long-lived during the simulation. Their life time can be made arbitrary long by using large
channels (cf. Table~\ref{table+model+adaptation}). 
We conclude that our microchannel system exhibits a nonequilibrium 
first order phase transition between two different migration
states. We find coexistence regions of the two states
(Fig.~\ref{fig+phase-transition+trajec200_combi} and
\ref{fig+phase-transition+split-mobil}). The nonmonotonic chain
length dependence of the mobility at $E = 0.04 E_0$ is a
consequence of that transition. 

\begin{figure}[t]
  \begin{center}
    \includegraphics[scale=0.3]{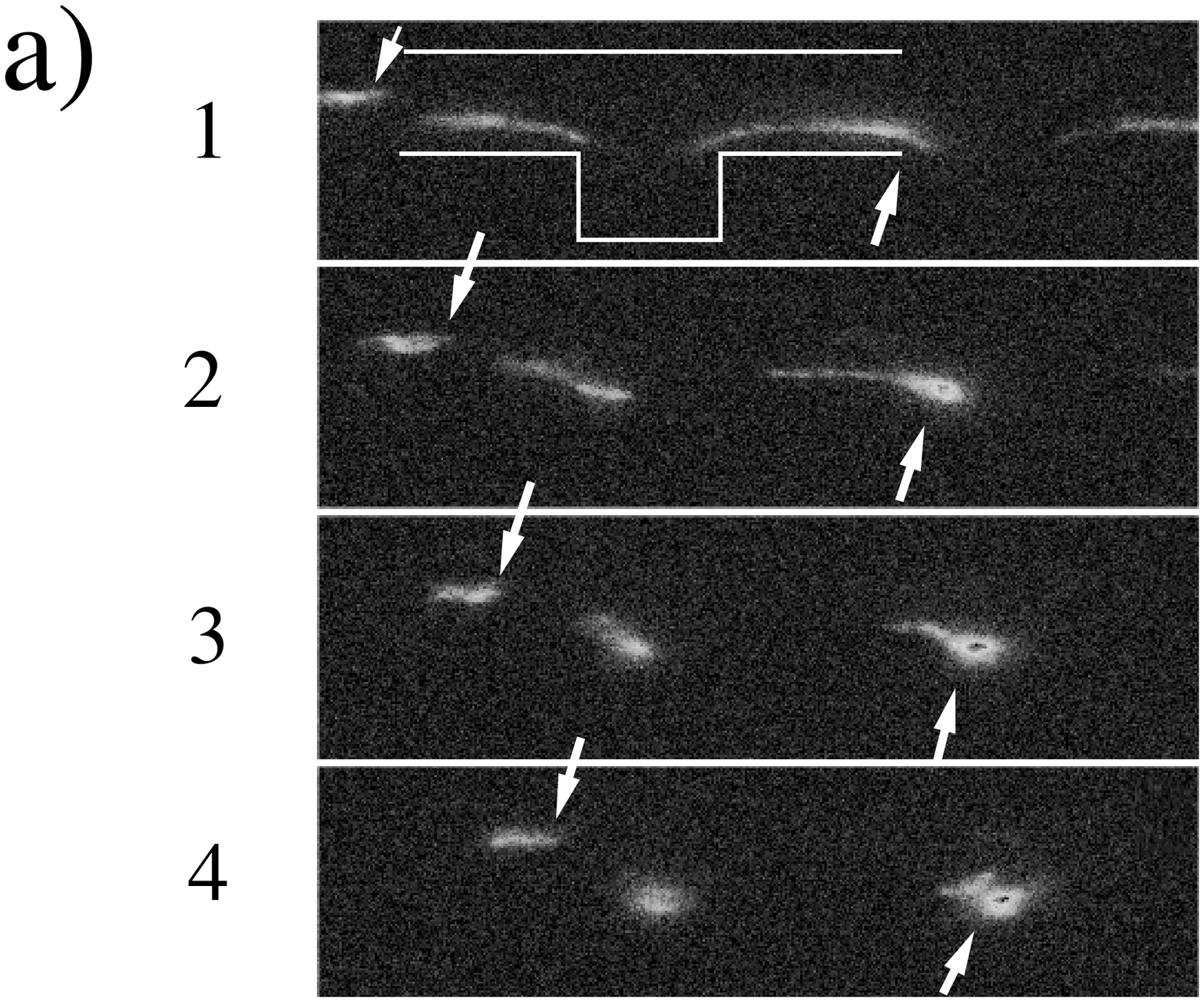}
    \vspace*{0.5cm}

    \includegraphics[scale=0.18]{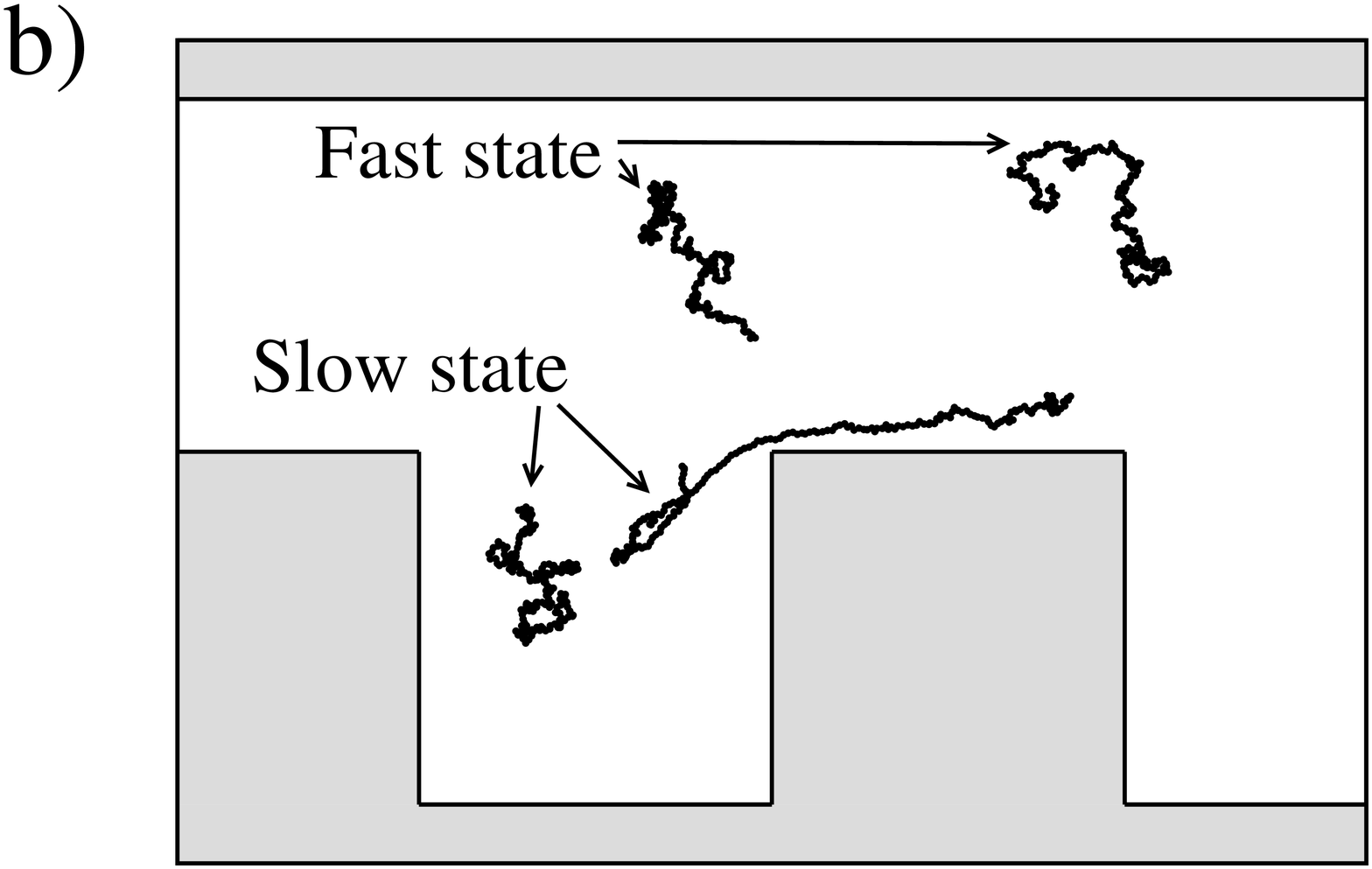}
  \end{center}
  \caption{Snapshots obtained from (a) an experiment with T2-DNA and
    (b) simulation with $200$ beads at $E = E_0$. The experimental
    snapshots show a series taken with 40 ms time step (fast state
    shown by arrows pointing down, slow state indicated by arrows
    pointing up). Both snapshots show chains in the slow state, which
    penetrate the wide region, form a coil inside and stretch  when
    passing into the narrow region. Chains in the fast state are also
    shown. Here the polymer remains permanently coiled in the
    homogeneous part of the field.}  
  \label{fig+phase-transition+snapshots}
\end{figure}

The experimental studies of $\lambda$-DNA and T2-DNA migrating in
large structures support our simulation results. In structures
characterized by $H \approx 5 \mu \mbox{m}$, at electric fields
ranging up to 100~V/cm, both $\lambda$- and T2-DNA exhibit two
distinct states of migration, which are illustrated in 
Fig.~\ref{fig+phase-transition+snapshots}a. In one state, the chain
remains coiled all the time in the upper (homogeneous) part of the
field, and travels at high speed. In the other, it penetrates deep
into each cavity, forms a coil inside, and gets stretched again when
moving into the narrow region. 


\begin{figure}[t]
  \begin{center}
    \includegraphics[scale=0.35,angle=0]{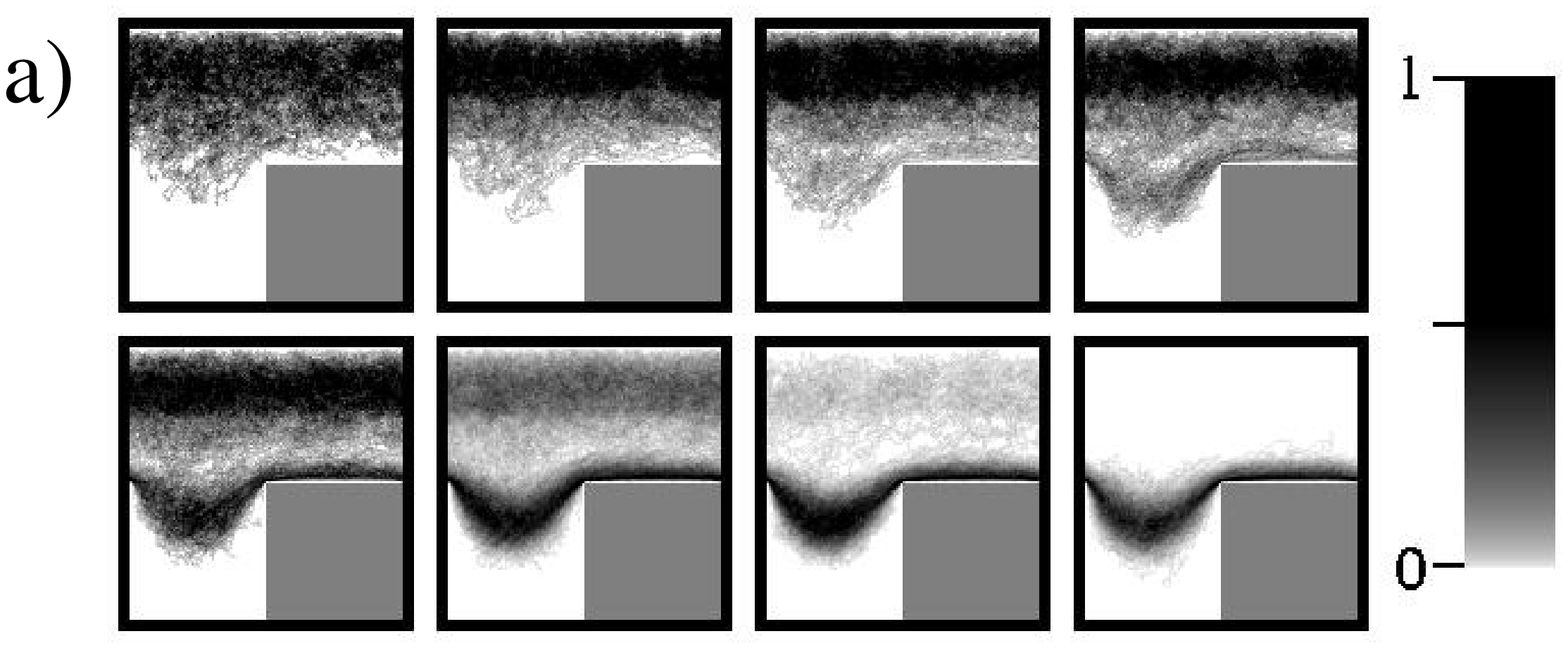}
    \hfill
    \includegraphics[height=3.cm,angle=0]{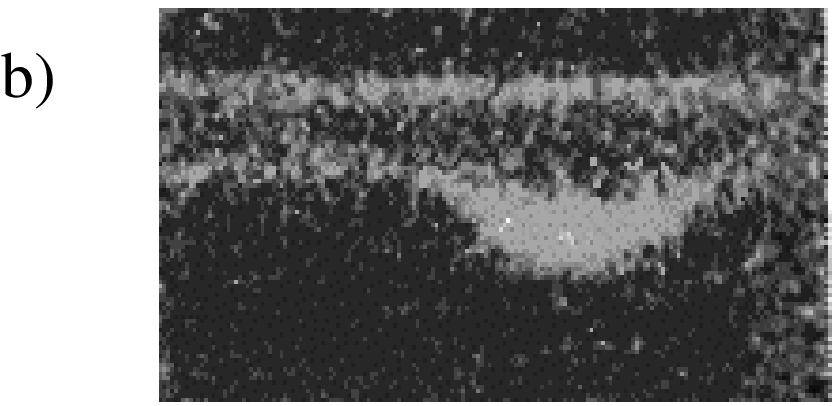}
  \end{center}
  \caption{(a)Monomer density histograms obtained from simulation for
    $N = 50, 100, 120, 140$ (top, left to right) and $N = 170, 200,
    250\; \mbox{and}\; 320$ (bottom) at $E = E_0$. The density
    distribution $\rho$ was cut off at $\rho = 1/2$ to dampen the
    peaks at the corners of the constrictions and to emphasize the
    depletion zone. (b) Monomer density histograms obtained from
    experiment at $E = 86 \mbox{V/cm}$ ($ 0.20 E_0$) for $\lambda$-DNA
    in $5\mu \mbox{m}$ constrictions. Note the depletion zone between
    both states.}  
  \label{fig+phase-transition+monomer-density}
\end{figure}

Fig.~\ref{fig+phase-transition+monomer-density} shows
monomer density histograms obtained from experiment and simulation.
In both the experiments and in the simulations with chains of length
$N = 120 - 250$ beads, we observe a low-density depletion zone between 
the upper and the lower part of the channel. Furthermore,
Fig.~\ref{fig+phase-transition+monomer-density}a shows that short 
chains are more likely to migrate in the fast state and long chains 
are more likely to migrate in the slow state. For $N = 120 - 250$,
the polymer migrates alternatingly in the  slow and fast state, 
as has already been seen from the trajectory of
$N = 200$ in Fig.~\ref{fig+phase-transition+trajec200_combi}. 

\begin{figure}[t]
  \begin{center}
    \includegraphics[scale=0.24,angle=0]{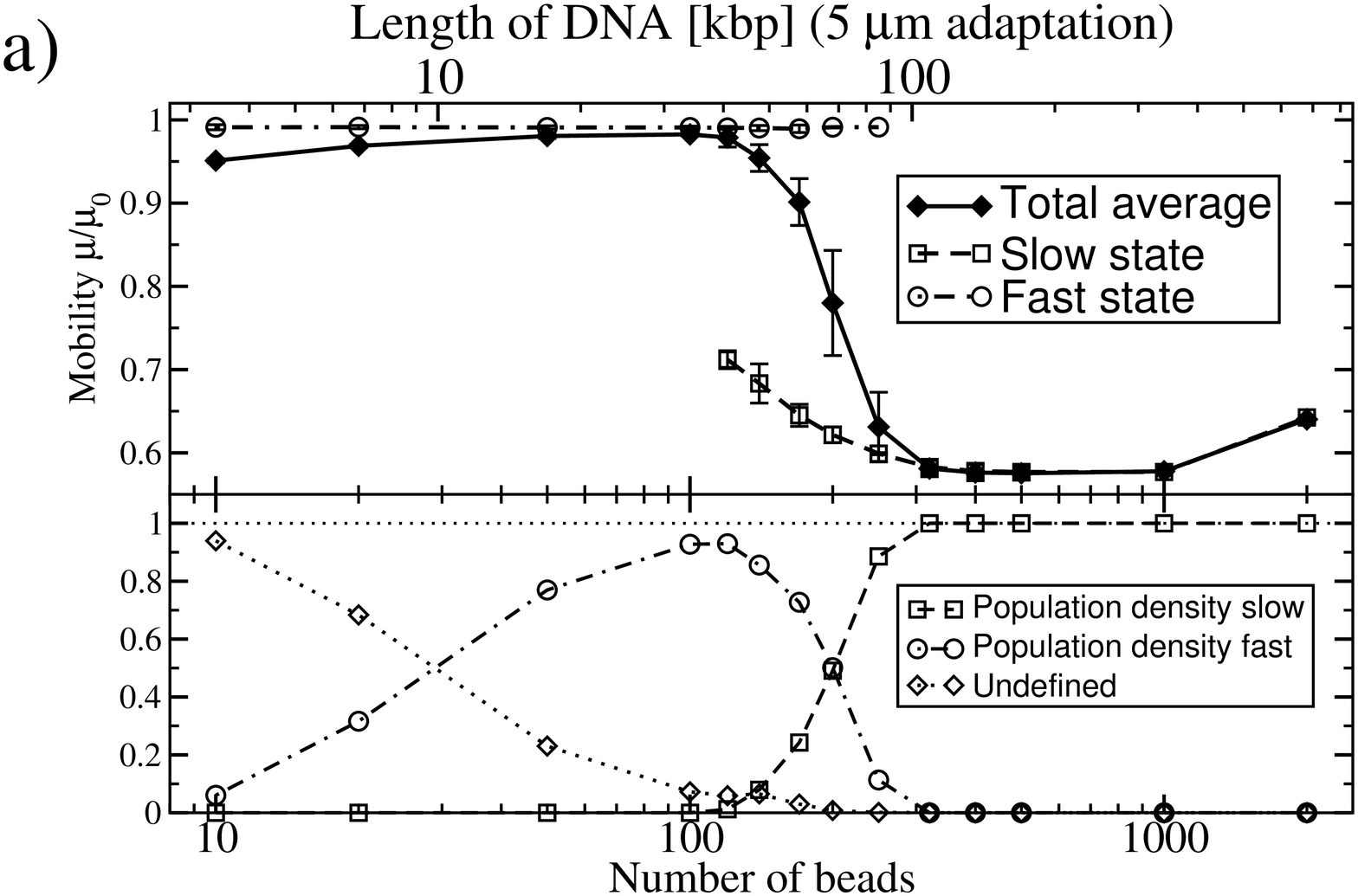}
    \includegraphics[scale=0.24,angle=0]{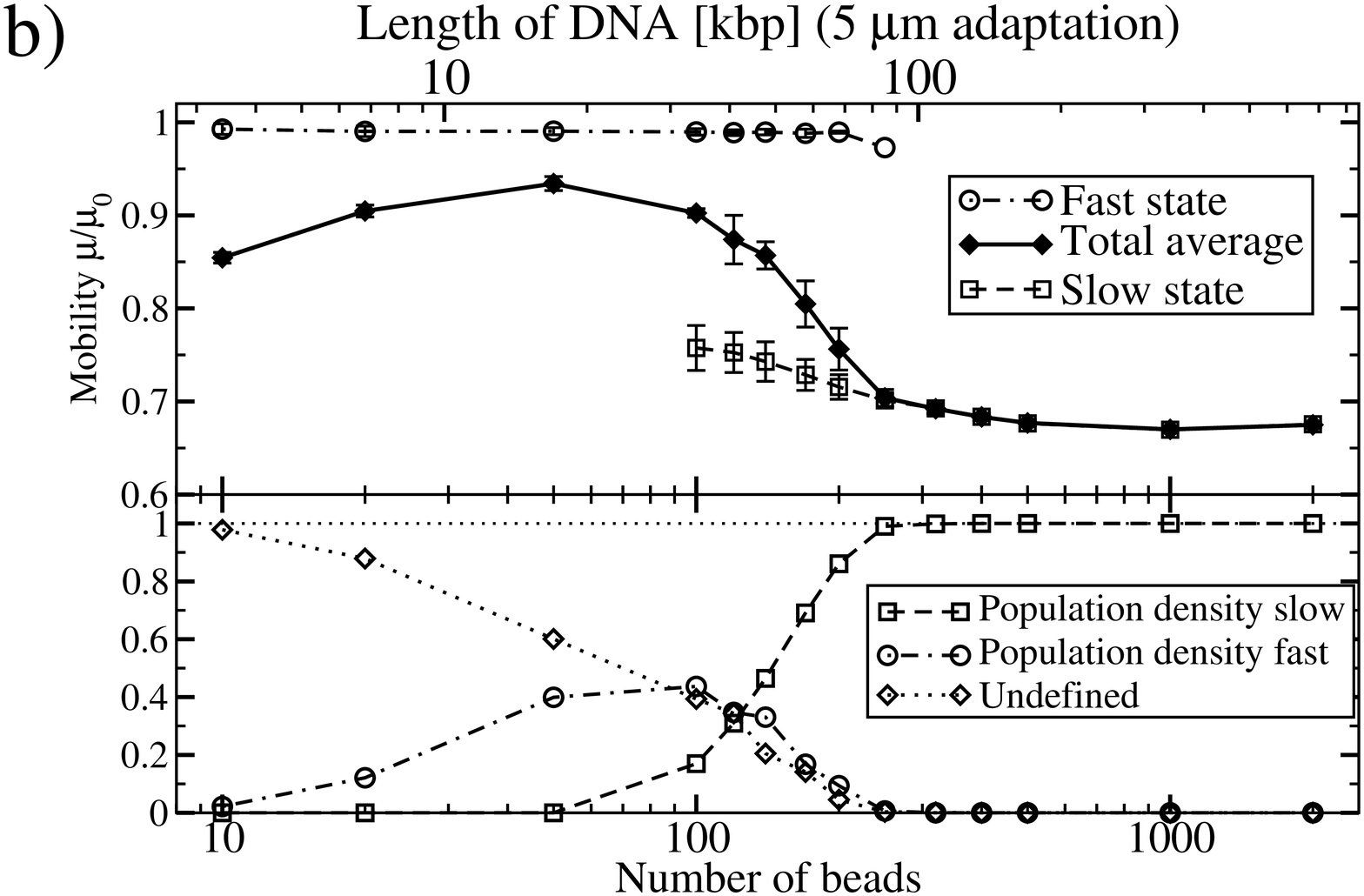}
    \includegraphics[scale=0.24,angle=0]{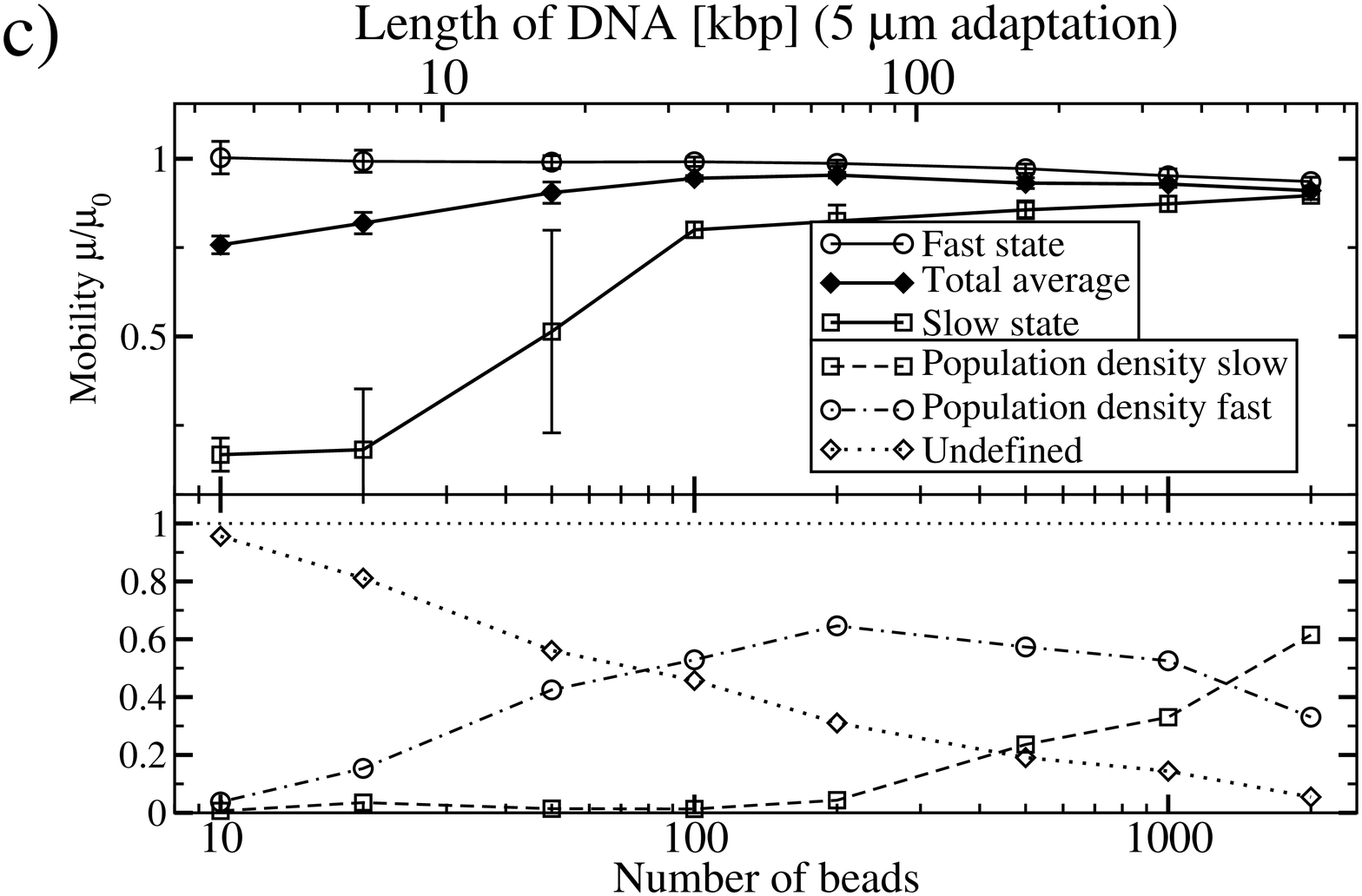}
  \end{center}
  \caption{Mobility as a function of the chains length at $E = 
    E_0$~(a), $E = 0.25 \: E_0$~(b) and $E = 0.04 \: E_0$~(c). Note the decrease in mobility at $N = 100 -
    250$ (34 kbp -- 85 kbp) at $E = E_0$, which is also reflected by the population densities given
    in the lower panel. For short chains, diffusion dominates the
    migration, and most of the time, it is not possible to assign
    unambiguously a single migration state. This is reflected by the
    amount of ``undefined'' population. The error bars of the slow
    and fast mobilities show the statistical error, based on the spread of the values. Note that the
    transition fades away when the electric field is decreased.} 
  \label{fig+phase-transition+split-mobil}
\end{figure}

%
%

Density histograms such as those shown in
Fig.~\ref{fig+phase-transition+monomer-density}a allow to determine
the population density of the two states. We choose the $z$-position
of the center of mass as an indicator of a polymer's state
(cf. also Fig.~\ref{fig+phase-transition+trajec200_combi}). Since the
transition does not take place instantaneously
(Fig.~\ref{fig+phase-transition+transition-fast-to-slow}), we define
two thresholds separated by a gap: a chain is taken to be in the slow
state or in the fast state, if $z \le 8 \sigma$ or $z \ge 20 \sigma$
over at least $5 \cdot 10^6$ time steps. After having assigned a state
to large portions of the trajectory, we can determine the mobility in
the slow and the fast state separately. 

Fig.~\ref{fig+phase-transition+split-mobil}a shows the results for
$E = E_0$ and compares it with the overall mobility. The assignment of
states works well for intermediate and long chains. For $N = 100$ (34
kbp), the polymer migrates almost completely in the fast state and for
polymers with $N \ge 320$ (110 kbp), the polymer occupies the slow
state only. Short chains diffuse very strongly in the $z$ direction,
no depletion zone occurs, and they often cross from the lower to the
upper part of the channel. Therefore, the assignment criterion often
fails, and only a few short fragments of the trajectory contribute to
the calculation of the mobility in the fast state. This explains why
the result does not coincide with the total mobility at low $N$. 

The transition was not only observed at $E = E_0$, but also observed
at smaller fields
$E$. Fig.~\ref{fig+phase-transition+split-mobil}b and Fig.~\ref{fig+phase-transition+split-mobil}c gives the 
mobilities and the population densities for $E = 0.25 \: E_0$ ($110
\mbox{V/cm}$) and $E = 0.04 \: E_0$ ($17 \mbox{V/cm}$), as determined
by the above criterium. At $E = 0.25 \: E_0$, the transition is not as
pronounced as for $E = E_0$, but still present. At $E = 0.04 \: E_0$,
the transition has almost disappeared, and only a slight decrease in
the overall mobility remains. This confirms our earlier assertion that
the nonmonotonic chain length 
dependence of the mobility at $E = 0.04 \: E_0$ is a signature of the
phase transition. It also clarifies why long chains (T2-DNA) migrate
slower than shorter chains ($\lambda$-DNA) in our channels. They tend
to penetrate deeper into the low field regions, which slows them down
significantly. 

The quantitative comparison between simulations and experiments raises
a question. According to the simulations, coexistence of two states
should be observed at chain lengths below $N=250$ beads, which
corresponds to 64 kbp in $5 \mu \mbox{m}$ microchannels. Thus we would
expect two mobilities for $\lambda$-DNA, but only one for
T2-DNA. Experimentally, we find that T2-DNA exhibits two states of
migration as well. This is in contrast to the prediction of our
simulation. However, a closer inspection reveals that even chains of
$N = 500$ beads (170 kbp, roughly T2-DNA) remain in a fast state for a
long time (up to $8 \cdot 10^8$ time steps, data not shown), if they
are prepared accordingly. These initial fast states turned out to be
so stable, that we were unable to analyze their life time in detail. 
The experimental situation hardly corresponds to an 
``equilibrated'' late-time limit.  It is conceivable that after 
being introduced into the microfluidic channel, a sizable fraction 
of molecules happens to be prepared in a state that is very weakly
populated in the long-time average.



\section{Migration state transition mechanisms}
\label{chapter+transition-mechanism}

To understand the transition and identify the
factors which stabilize the two migration states, we
have investigated the transition processes between
the two states in more detail.


\subsection{Transition from the fast to the slow state}
\label{chapter+phase-transition+fast-to-slow}

\begin{figure}[b]
  \begin{center}
    \includegraphics[scale=0.28,angle=0]{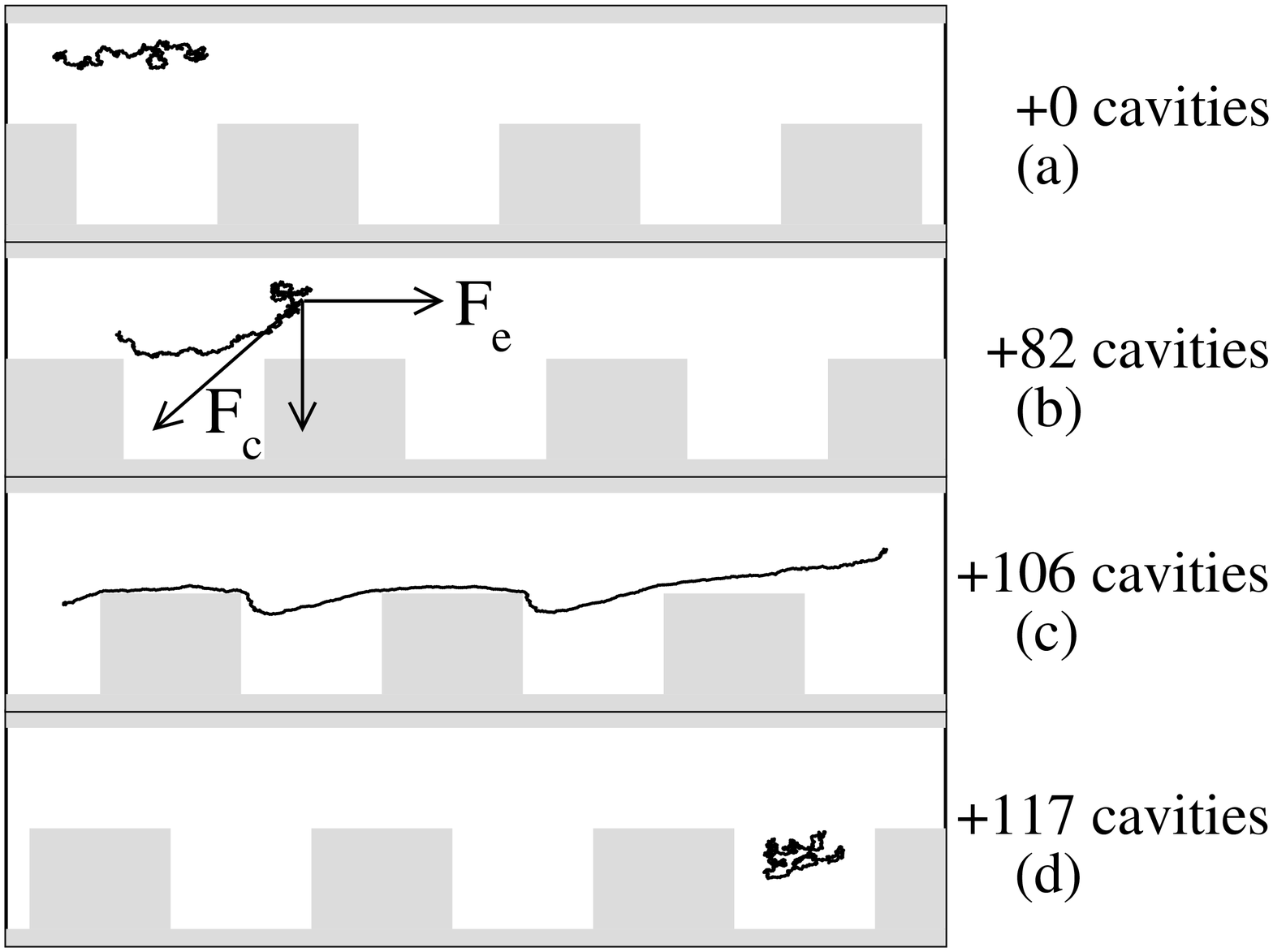}
  \end{center}
  \caption{Transition of chains with $N = 500$ beads (170 kbp) from
    the fast to the slow state. The snapshots are shifted by the
    number of cavities given on the right. The transition itself takes
    about $35$ cavities. In b) a force parallelogram is sketched
    schematically. The electric force $F_{\mbox{\tiny e}}$ and the
    entropic spring force $F_{\mbox{\tiny c}}$ on the leading monomers
    add up to a total force which has a net downward component. } 
  \label{fig+phase-transition+transition-fast-to-slow}
\end{figure}

We begin with discussing the transition from the fast to the slow
state. A series of snapshots of a chain with $N=500$ beads in the
process of crossing from the fast to the slow state is shown in
Fig.~\ref{fig+phase-transition+transition-fast-to-slow}. The chain
travels through roughly 35 cavities during the transition process,
which we can divide into five stages:

\begin{enumerate}
\item At the beginning, the polymer is in a coiled conformation in the
  homogeneous field region. The radius of gyration of a chain of $N =
  500$ beads in free solution is $R_{\mbox{\tiny g}} \approx 18
  \sigma$. The size of the homogeneous field region is limited, thus
  larger polymers are deformed
  (Fig.~\ref{fig+phase-transition+transition-fast-to-slow}a). 
\item At some stage, one polymer loop or end happens to penetrate into
  a region with a lower field. As that polymer part experiences a
  lower pulling force, it slows down, while the force acting on the
  rest of the polymer is unchanged. This leads to uncoiling and
  stretching of the polymer. 
\item More and more monomers are slowed down. Since the field in the
  homogeneous part pulls horizontally in the $x$-direction, and the
  delayed monomers are located in the cavities at low $z$ values, the
  total force acting on the chain has a net downward component 
  (Fig.~\ref{fig+phase-transition+transition-fast-to-slow} b). 
  After a while, the polymer is completely stretched 
  (Fig.~\ref{fig+phase-transition+transition-fast-to-slow} c). 
\item For the highly stretched chain, 
  (Fig.~\ref{fig+phase-transition+transition-fast-to-slow} c), 
  the probability that the foremost monomer gets trapped in a cavity
  is very high. Immediately afterwards, the polymer collapses in that
  cavity. (Fig.~\ref{fig+phase-transition+transition-fast-to-slow}
  d). 
\item At the end, the polymer migrates in the slow state. For long
  polymers, the probability to leave the slow state is very low
  (Fig.~\ref{fig+phase-transition+transition-fast-to-slow} d).  
\end{enumerate}



\subsection{Transition from the slow to the fast state}
\label{chapter+phase-transition+slow-to-fast}

Unfortunately, the configuration snapshots of the transition from the
slow to the fast state cannot be interpreted as easily as those in 
Fig.~\ref{fig+phase-transition+transition-fast-to-slow}, and we have
not yet been able to distil the mechanisms which stabilize the fast
state in a satisfactory way. 

One effect which presumably favors the transtion is the inertia of the
molecule at the corners of the constriction. Monomers migrating up at
high speed against the wall continue to drift upwards slightly as they
enter the constriction (data not shown). This effect involves
dynamical processes on time scales smaller than $\tau_0$, the
characteristic decay time for the drift velocity of the polymer. In
our model, $\tau_0$ is given by $\tau_0 = m/\zeta$, which corresponds
to approximately  $10^{-4} s$ for beads of mass $m = \zeta t_0$. For
DNA in free solution, the drift velocity decay time
is~\cite{Grossmann92} $\tau_0 \approx 10^{-9} \dots
10^{-12}\mbox{s}$, which is much smaller. Thus this particular
drift effect is an artifact of the simulation model. 
To verify that this mechanism is not responsible for the transition
observed in the simulations, we modified the model by changing its
dynamics to that of an overdamped system~\cite{Reimann2002}
(Eq.~\ref{eqn+model+langevin-no-verlet-dynamics}). We could only
simulate short chains, and were unable to perform a thorough analysis
due to the reduced time
step~(Eq.~\ref{eqn+model+langevin-no-verlet-dynamics}). The transition
still occurs in both directions (data not shown). This proves that the
fast state is stabilized by another mechanism. In fact, the only
noticeable effect of inertia was to {\em suppress} the transition for
short chains.  

Another conceivable mechanism which might favor and stabilize the 
fast phase is based on diffusion. While the polymer migrates along the
wall in the narrow region and around the corners, it can only diffuse
upwards. This effect is enhanced by the tendency of the polymer, which
has been forced into a stretched conformation in the inhomogeneous
parts of the field, to contract back to a coil. The characteristic
time scale for contraction is presumably related to the Rouse time and
scales like $N^{1+2 \nu}$~\cite{Doi86}, which
would explain why the slow state dominates at large chain length. We
have not yet been able to support this explanation quantitatively. 

The understanding of the fast phase remains an open
problem. Systematic studies are difficult, because the opposite
transition is triggered by diffusion as well. 



\subsection{Critical chain length}
\label{chapter+phase-transition+crossover-point}

Fig.~\ref{fig+phase-transition+transition-fast-to-slow} suggests that
the transition can only take place if the electric field exceeds a
critical value; the inhomogeneous field has to uncoil the chain and
thus work against the entropic spring force of the polymers. The mean
elongation $\langle r \rangle$ of stretched self-avoiding
polymers 
subject to a force $f$, which pulls on the end monomers,
is~\cite{Baumgaertner2002}
\begin{equation}
  \langle r \rangle \propto N a \left( f a / k_{\mbox{\tiny B}} T
  \right)^{2/3}, 
  \label{eqn+phase-transition+strong-stretching}
\end{equation}
for strongly stretched chains, and
\begin{equation}
  \langle r \rangle \propto {a^2 N^{2\nu}}/ (k_{\mbox{\tiny B}} T) f
  \label{eqn+phase-transition+weak-stretching}
\end{equation}
for weakly stretched chains, 
where $a \approx 0.5 \sigma $ is the proportionality constant in
$R_{\mbox{\tiny g}} = a N^\nu$. In our case, the force of course acts
on all monomers. Nevertheless, we shall use
Eq.~\ref{eqn+phase-transition+strong-stretching} and Eq.~\ref{eqn+phase-transition+weak-stretching} to roughly 
estimate the maximal effect of the inhomogeneous field on a chain, with $f \propto E$. 

The question is: how much does the field have to stretch a chain in
order to stabilize a distinct slow state? Let us consider the critical
chain length $N_{\mbox{\tiny c}}$ at which both states are equally
populated. Assuming that it is sufficient to stretch the chains by an
amount proportional to their own size, one would expect $N E^{5/3}$ to
be a constant at $N = N_{\mbox{\tiny c}}$ (according to both Eq.~\ref{eqn+phase-transition+strong-stretching} and
Eq.~\ref{eqn+phase-transition+weak-stretching}). Alternatively, it 
might be necessary to stretch the chain to a fixed length, which is
determined by the size of the device, in particular the constriction
length $H$ (see Fig.~\ref{fig+introduction+channel}b). In that case, according to
Eq.~\ref{eqn+phase-transition+strong-stretching}, $N E^{2/3}$ should be constant. 
\begin{table}[t]
  \begin{tabular}{|r||r|r|r|} \hline
    $E / E_0$ & $N_{\mbox{\tiny c}}$ & $N_{\mbox{\tiny c}} \cdot (E/E_0)^{5/3}$
    & $N_{\mbox{\tiny c}} \cdot (E/E_0)^{2/3}$ 
    \\\hline\hline 
    $0.08$ & $290 \pm 25$ & $4.3 \pm 0.4$   & $55 \pm 5$  \\\hline
    $0.15$ & $174 \pm 20$ & $7.4 \pm 0.9$   & $50 \pm 6$  \\\hline
    $0.25$ & $124 \pm 15$ & $12.3 \pm 1.5$  & $50 \pm 6$  \\\hline
    $0.35$ & $120 \pm 15$ & $20.9 \pm 2.6$  & $60 \pm 8$  \\\hline
    $0.50$ & $106 \pm 10$ & $33.4 \pm 3.2$  & $67 \pm 6$  \\\hline
    $0.70$ & $125 \pm 10$ & $71.7 \pm 5.7$  & $103 \pm 8$ \\\hline
    $1.00$ & $200 \pm  5$ & $200.0 \pm 5.0$ & $200 \pm 5$ \\\hline
  \end{tabular}
  \caption{Crossover lengths $N_{\mbox{\tiny c}}$ for various electric fields. Errors of
    the crossover lengths are estimates of the population density
    histograms.} 
  \label{table+phase-transition+crossover}
\end{table}
This is tested in table~\ref{table+phase-transition+crossover}. The
quantity $N_{\mbox{\tiny c}} \cdot E^{5/3}$ is obviously not
constant for different electric fields $E$. However, 
$N_{\mbox{\tiny c}} \cdot E^{2/3}$ is almost constant for $E \le 0.50
E_0$. For larger values of $E$ this rule breaks down. This
might be related to the inertia effects discussed in
Sec.~\ref{chapter+phase-transition+slow-to-fast}. Simulations
with overdamped dynamics at $E = E_0$ indeed suggest that, as $m \to
0$, chains as small as $N = 50$ already exhibit two distinct
migration speeds (data not shown). This would yield $N_{\mbox{\tiny
    c}}E^{2/3} = 50 E_0^{2/3}$, which is consistent with results at
lower fields. Unfortunately, the statistical quality of our data is
not sufficient to determine $N_{\mbox{\tiny c}}$ rigorously in the
overdamped case. 

Nevertheless, the sum of our findings suggests that the critical
length is determined by the geometry of the device rather than by the
size of the chain. This could explain why the transition disappears
at small fields: for $E = 0.04 E_0$, the predicted critical length is
$N_{\mbox{\tiny c}} \approx 500$, but free chains of that length have
an end-to-end distance of $46\sigma$, which is comparable to the
channel height $H$. Hence a fast state, in which the chain is
essentially unperturbed, cannot exist.

This result suggests strategies for the design of microstructures 
which do or do not exhibit the two-state behavior; we expect it can be
suppressed by choosing the constriction length $d$ in Fig.~\ref{fig+introduction+channel}b small,
and it can be promoted by making $d$ larger.




\section{Conclusion}
\label{chapter+conclusions}

We have presented Brownian dynamics simulations and
experimental studies of DNA migration in structured microchannels. Our
simulations reproduce earlier experimental results presented by Duong
et al.~\cite{Duong2003} and, in particular, explain the experimentally
observed migration order of $\lambda$- and T2-DNA in migration studies
and separation experiments~\cite{Duong2003}: in channels with
geometries as shown in Fig.~\ref{fig+introduction+channel}, and for
moderate field values, the shorter $\lambda$-DNA molecules migrate
faster than longer T2-DNA molecules. 

This behavior, which is opposite to that expected at weaker fields,
is a signature of a high-field non-equilibrium phase transition. At
very high fields (or in much larger structures), we found that chains can
migrate with two distinctly different speeds, assuming different
states of migration. This was observed both experimentally and in
simulations. We have discussed the factors which stabilize the two
states and can thus propose strategies for the design of channels which 
do or do not exhibit the transition, and which could be suitable for separation 
experiments. 

The comparison of simulations and experiments demonstrates that our
simple DNA model, which disregards electrostatic interactions, 
hydrodynamic interactions, and irrotational fluid flow,
nevertheless reproduce the experimental results of DNA migration in our
microchannels. However, the interplay of DNA motion and buffer flow 
is non-trivial and may be neglected for irrotational EOF outside Debye layers only. 
As mentioned in Sec.~\ref{chapter+model+adaptation}, the experimental electroosmotic
mobility in straight channels is of the same order of magnitude as the
DNA mobility. In structured microchannels, where DNA molecules migrate at either high fields or along the
Debye layer, electroosmosis should give
rise to interesting nontrivial flow patterns which certainly influence
the DNA migration. Moreover, the electrostatic and the hydrodynamic
interactions are not entirely screened in microchannels, which leads
to additional effects~\cite{Long96}. In order to investigate such
phenomena, systematic experimental studies are necessary, and
efficient new simulation methods need to be developed. These shall be
explored in the future. 

From a practical point of view, our results have two important implications:

On the one hand, they demonstrate that the physics of DNA electrophoresis in microchannels is surprisingly
complex. When designing new microfluidic devices, one must be aware of the possibility of nonmonotonic or even
bistable behavior, and a thorough theoretical analysis is advisable.

On the other hand, the observed two-state behavior can presumably be exploited. Our findings indicate that the
migration velocities of the DNA fragments differ by a large factor (almost $2$ in
Fig.~\ref{fig+phase-transition+split-mobil}) for $N > N_{\mbox{\tiny c}}$ and $N < N_{\mbox{\tiny c}}$. Since
the chain length corresponding to $N_{\mbox{\tiny c}}$ depends on the channel geometry
(cf. Table~\ref{table+model+adaptation}), a device with a gradient in the constriction length could in principle
sequentially separate molecules very efficiently. However, as the switching time between the two states is
very long, methods need to be developed to accelerate the transition. We are currently exploring corresponding strategies experimentally as well as theoretically. 



%



\section*{Acknowledgements}

We would like to thank Ralf Eichhorn for carefully reading the
manuscript and the referees for their useful comments. This work was funded by the German Science Foundation
(SFB 613, Teilprojekt D2). Parts of the simulation jobs were handled by the
job queuing system Condor, which was developed by the Condor Team at
the Computer Science Department of the University of
Wisconsin~\cite{Condor}.  


\bibliography{electrophoresis}

\end{document}